\documentclass[10pt,a4paper]{article}

\usepackage{natbib}
\pdfoutput=1

\usepackage{graphicx}

\usepackage{caption}
\usepackage{subcaption}

\usepackage{multirow}

\usepackage{amsmath,amssymb}

\usepackage{authblk}

\title{Mortality rate forecasting: can recurrent neural networks beat the Lee-Carter model?}
\author[1]{G\'abor Petneh\'azi\thanks{gabor.petnehazi@science.unideb.hu}}
\author[2]{J\'ozsef G\'all\thanks{gall.jozsef@inf.unideb.hu}}
\affil[1]{Doctoral School of Mathematical and Computational Sciences, University of Debrecen}
\affil[2]{Department of Applied Mathematics and Probability Theory, University of Debrecen}
\date{}

\begin{document}
\maketitle

\begin{abstract}
This article applies a long short-term memory recurrent neural network to mortality rate forecasting. The model can be trained jointly on the mortality rate history of different countries, ages, and sexes. The RNN-based method seems to outperform the popular Lee-Carter model.
\end{abstract}

\section{Introduction}
Human mortality rates form a particularly challenging task for time series forecasting. Forecasts should be produced separately for different ages, preferably for multiple years ahead into the future, having just a relatively small amount of historical data available. It is pretty difficult to create and evaluate forecasts under such circumstances.\\
In this paper, we apply a recurrent neural network to mortality rate forecasting. RNNs are usually used in data rich environments. Short time series (like the registered history of human mortality rates) do not hold much promise for such flexible machine learning models. However, we argue that when applied properly, neural networks might outperform well proven mortality forecasting algorithms.\\
In order to feed as much data to the neural networks as possible, we build joint machine learning models that can learn from multiple mortality rate series.\\
We apply a long short-term memory network and compare its forecasting performance to that of the deservedly popular Lee-Carter method. We evaluate the algorithms on the last 10 years of the latest life tables of countries around the world.

\section{Literature Review}
A few recent studies applied neural networks to mortality rate forecasting, but this area still seems rather unexplored and unexploited. \citet{shah2009mortality} applied neural networks to cause-specific mortality data, and found it to perform close to or better than the Lee-Carter model, though they were difficult to compare in that context. \citet{hainaut2018neural} used a neural network based generalization of principal component analysis to summarize mortality information as part of a forecasting framework. \citet{richman2018neural} extended the Lee-Carter model to multiple populations using neural networks, and achieved competitive out-of-sample forecasting performance. \citet{werpachowska2018forecasting} used recurrent neural networks to forecast UK mortality rates.
\citet{nigri2019deep} used LSTM to forecast the time index of the Lee-Carter model.

\section{Lee-Carter model}
\citet{lee1992modeling} proposed a method to make long-run forecasts of age-specific mortality rates. It has since become the leading statistical model of mortality \citep{deaton2004mortality}. The model is widely applied with several generalizations and extensions. We use a basic version of the Lee-Carter model as a baseline for neural networks in our study.\\
Let $m_{x,t}$ denote the mortality rate of age group $x$ in year $t$. The Lee-Carter model makes an estimate of the log-mortality rates (\ref{eq:lc_eq}).\\
\begin{equation} \label{eq:lc_eq}
ln(m_{x,t}) = a_x + b_xk_t + e_{x,t}
\end{equation}
$a_x$ and $b_x$ are age-specific values, while $k_t$ varies in time. $a_x$ is the general shape of mortality across age, and $b_x$ shows the rate of change for different ages with respect to deviations in $k_t$. $k_t$ can be modeled by statistical time series methods, which empowers the model to make forecasts into the future.\\
In order to fit the model, we could use the least squares method, minimizing (\ref{eq:lc_loss}).
\begin{equation} \label{eq:lc_loss}
\sum_{x} \sum_{t}  (ln(m_{x,t})-a_x-b_xk_t)^2
\end{equation}
Though it is undetermined, so we need conditions (\ref{eq:lc_sumb}) and (\ref{eq:lc_sumk}) to hold in order to have a unique solution.
\begin{equation} \label{eq:lc_sumb}
\sum_{x} b_x = 1
\end{equation}
\begin{equation} \label{eq:lc_sumk}
\sum_{t} k_t = 0
\end{equation}
The estimates of $a_x$ are just the average values of $ln(m_{x,t})$ over time (\ref{eq:lc_ax}).
\begin{equation} \label{eq:lc_ax}
\hat{a}_x = \frac{1}{T} \sum_{t} ln(m_{x,t})
\end{equation}
The estimates of $b_x$ and $k_t$ are obtained from the singular value decomposition of the centered log-mortality matrix (\ref{eq:lc_mxt}).
\begin{equation} \label{eq:lc_mxt}
M_{x,t} = ln(m_{x,t}) - \hat{a}
\end{equation}
The estimates using the SVD $M = U D V^T$ are given by (\ref{eq:lc_bx}) and (\ref{eq:lc_kt}).
\begin{equation} \label{eq:lc_bx}
\hat{b}_x = \frac{U_{x,1}}{\sum_x U_{x,1}}
\end{equation}
\begin{equation} \label{eq:lc_kt}
\hat{k}_t = V_{t,1} D_{1,1} \sum_x U_{x,1} 
\end{equation}
Lee and Carter suggested a second stage estimation of $k_t$ so that the predicted number of deaths equals the observed number of deaths. However, this step is not self-evident, several different solutions have been proposed. For the sake of simplicity, we omit the second stage estimation in this study.\\
The $k_t$ values form a time series which may be modeled and forecasted using the Box-Jenkins method. In the original article, Lee and Carter proposed a random walk with drift, that is, an ARIMA(0,1,0). We use the same model.

\section{Long short-term memory}
LSTM \citep{hochreiter1997long} is a gated memory unit for recurrent neural networks. It has 3 gates (sigmoid functions) that manage the memory content, enabling the system to read, write and forget data, and thus to have a theoretically infinite memory. Long short-term memory might be described by 5 simple equations (\ref{eq:lstm_eqns}).\\
\begin{equation} \label{eq:lstm_eqns}
\begin{split}
i_{t} = sigmoid \left ( W_{i}x_{t} + U_{i}h_{t-1} + b_{i} \right )\\
f_{t} = sigmoid \left ( W_{f}x_{t} + U_{f}h_{t-1} + b_{f} \right )\\
o_{t} = sigmoid \left ( W_{o}x_{t} + U_{o}h_{t-1} + b_{o} \right )\\
c_{t} = f_{t} \odot c_{t-1} + i_{t} \odot tanh \left ( W_{c}x_{t} + U_{c}h_{t-1} + b_{c} \right )\\
h_{t} = o_{t} \odot tanh \left ( c_{t} \right )
\end{split}
\end{equation}
$x_{t} \in \mathbb{R}^n$ is the input, $h_{t} \in \mathbb{R}^h$ is the output (or hidden state) of the LSTM. $c_{t} \in \mathbb{R}^h$ is the the memory (or cell state). $i_{t} \in \mathbb{R}^h$, $f_{t} \in \mathbb{R}^h$, and $o_{t} \in \mathbb{R}^h$ are the input gate, forget gate and output gate, respectively. $W \in \mathbb{R}^{h \times n}$, $U \in \mathbb{R}^{h \times h}$ and $b \in \mathbb{R}^h$ are trainable weights.\\ 
All the weights might be optimized using backpropagation---the LSTM units have a self-managing memory. They learn what to remember, they learn what to forget. It makes them a popular choice for various sequence learning tasks.\\
LSTM networks are often applied to text data (e.g. machine translation) with amazing performance. However, time series forecasting is a tough task due to the scarcity of data. Even more so when we only have yearly observations.\\
We have much better chances when we can use multiple time series to learn the same process. The case of mortality rate forecasting belongs here. We have separate observations for all age groups, which we might expect to follow similar patterns. The Lee-Carter model assumes that there is a common mortality trend $k_t$ which can help forecast the mortality of either age group. In a similar vein, we apply LSTM networks to learn and memorize the patterns of mortality rate changes of all age groups, and use the common knowledge to make separate forecasts.\\
We might even use multiple time series datasets to learn a single general mortality rate forecaster. For example, we can use multiple countries' data to build a joint machine learning model that can make forecasts for any one of the countries. Or, similarly, we could learn from female and male mortality rates and apply it to forecast the totals.\\
We aimed to find reasonable hyperparameters for the LSTM network, though we did not use any systematic procedure, just trials and intuition. We ended up with a small and simple model architecture. The joint models could probably have worked better with more parameters and thus higher complexity.\\
The 10-year forecasts were produced in a recursive manner: single-year forecasts were iteratively fed back to the algorithm.\\
We used a one-layer LSTM with 8 units, followed by a dense layer with a single unit with linear activation. The network was unrolled to 16 steps, and the mortality rates were fed to the algorithm in 128-item batches. The data was standardized by subtracting the mean and dividing by the standard deviation of the whole training set. We chose the Adam optimizer \citep{kingma2014adam} to minimize the mean squared error loss function. The learning rate was set to 0.001, and the networks were trained for 300 epochs.\\

\section{Data}
Our datasets were obtained from mortality.org \citep{hmd}.\\
We have downloaded the life tables of all 40 countries available in the database. The length of registered mortality rate history varies from country to country. We had to exclude 4 countries with too few observations, and 1 country with missing values.\\
We are forecasting the mortality rates (or death rates), that is, the proportion of deaths in the given age group, in the given time.

\section{Evaluation}
We make multi-step out-of-sample forecasts for the last 10 years of the available mortality data. The validation time periods differ from country to country, since the datasets might have been updated more or less recently.\\
We use different evaluation metrics, so that the comparison is as complete as possible. RMSE (\ref{eq:eval_rmse}) is the square root of mean squared error. The LSTM method uses the mean squared error loss function, so this metric is directly optimized during the training. MAE (\ref{eq:eval_mae}) is the mean, while MedAE (\ref{eq:eval_medae}) is the median of absolute errors---they might be less sensitive to outliers, and maybe also easier to interpret. SMAPE (\ref{eq:eval_smape}) (symmetric mean absolute percentage error) is the mean of absolute errors divided by half the sum of the corresponding actual and forecasted (absolute) values, expressed as a percentage. It differs from the other metrics in that it is a relative measure. ME (\ref{eq:eval_me}) is simply the mean of the errors, when it is other than 0, it might indicate bias in one direction.
\begin{equation} \label{eq:eval_rmse}
RMSE(y, \hat{y}) = \sqrt{\frac{1}{n}\sum_{i=1}^{n}(y_i - \hat{y}_i)^2}
\end{equation}
\begin{equation} \label{eq:eval_mae}
MAE(y, \hat{y}) = \frac{1}{n} \sum_{i=1}^{n}|y_i - \hat{y}_i|
\end{equation}
\begin{equation} \label{eq:eval_medae}
MedAE(y, \hat{y}) = median(|y_1 - \hat{y}_1|, \dots , |y_n - \hat{y}_n|)
\end{equation}
\begin{equation} \label{eq:eval_smape}
SMAPE(y, \hat{y}) = \frac{100}{n} \sum_{i=1}^{n}\frac{|y_i - \hat{y}_i|}{([y_i] + |\hat{y}_i|)/2}
\end{equation}
\begin{equation} \label{eq:eval_me}
ME(y, \hat{y}) = \frac{1}{n} \sum_{i=1}^{n}(\hat{y}_i - y_i)
\end{equation}

\section{Experiments and Results}

\begin{figure}
\centering
\begin{subfigure}{.5\textwidth}
  \centering
  \includegraphics[width=1.\linewidth]{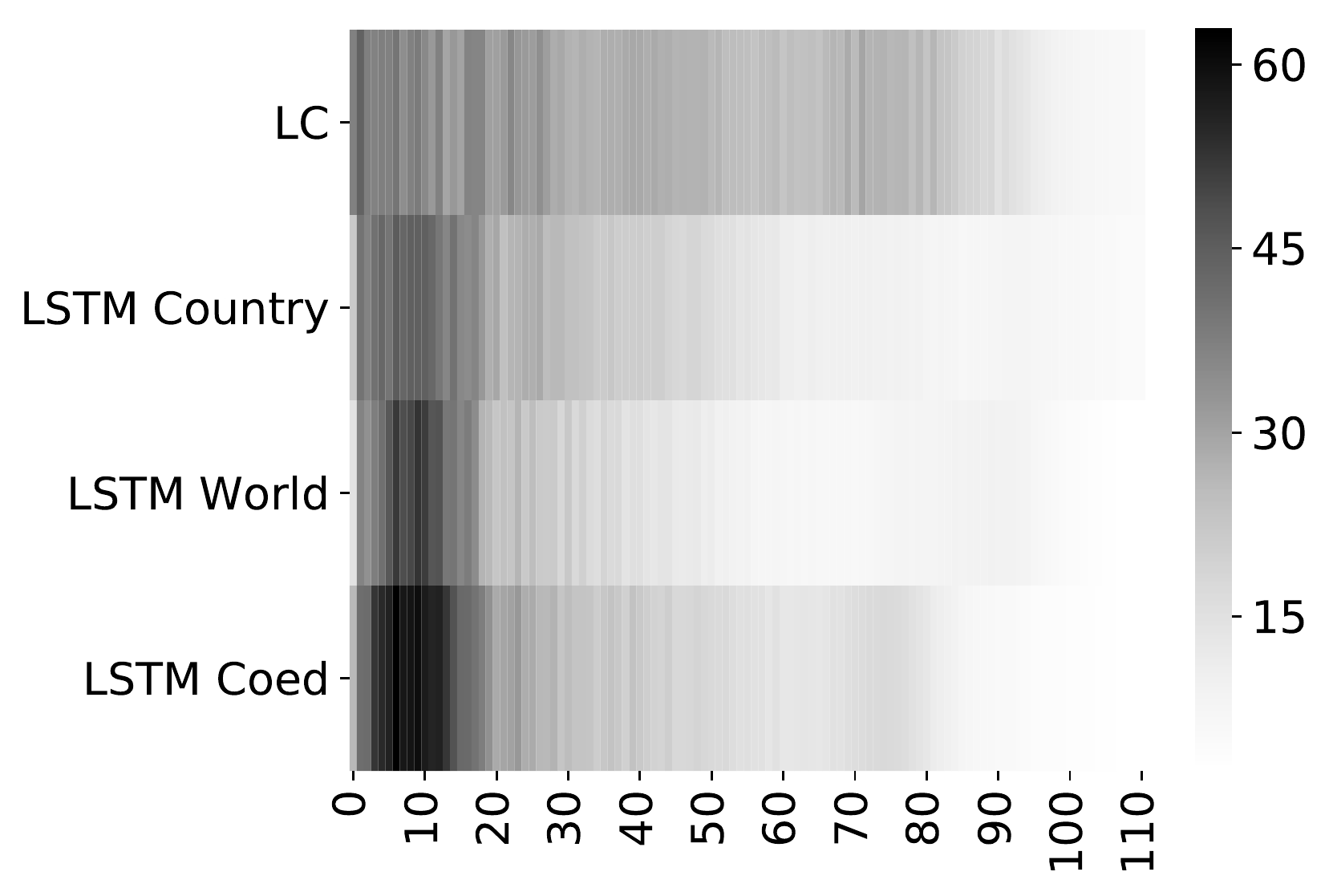}
  \caption{SMAPE}
  \label{fig:errors_age_smape}
\end{subfigure}%
\begin{subfigure}{.5\textwidth}
  \centering
  \includegraphics[width=1.\linewidth]{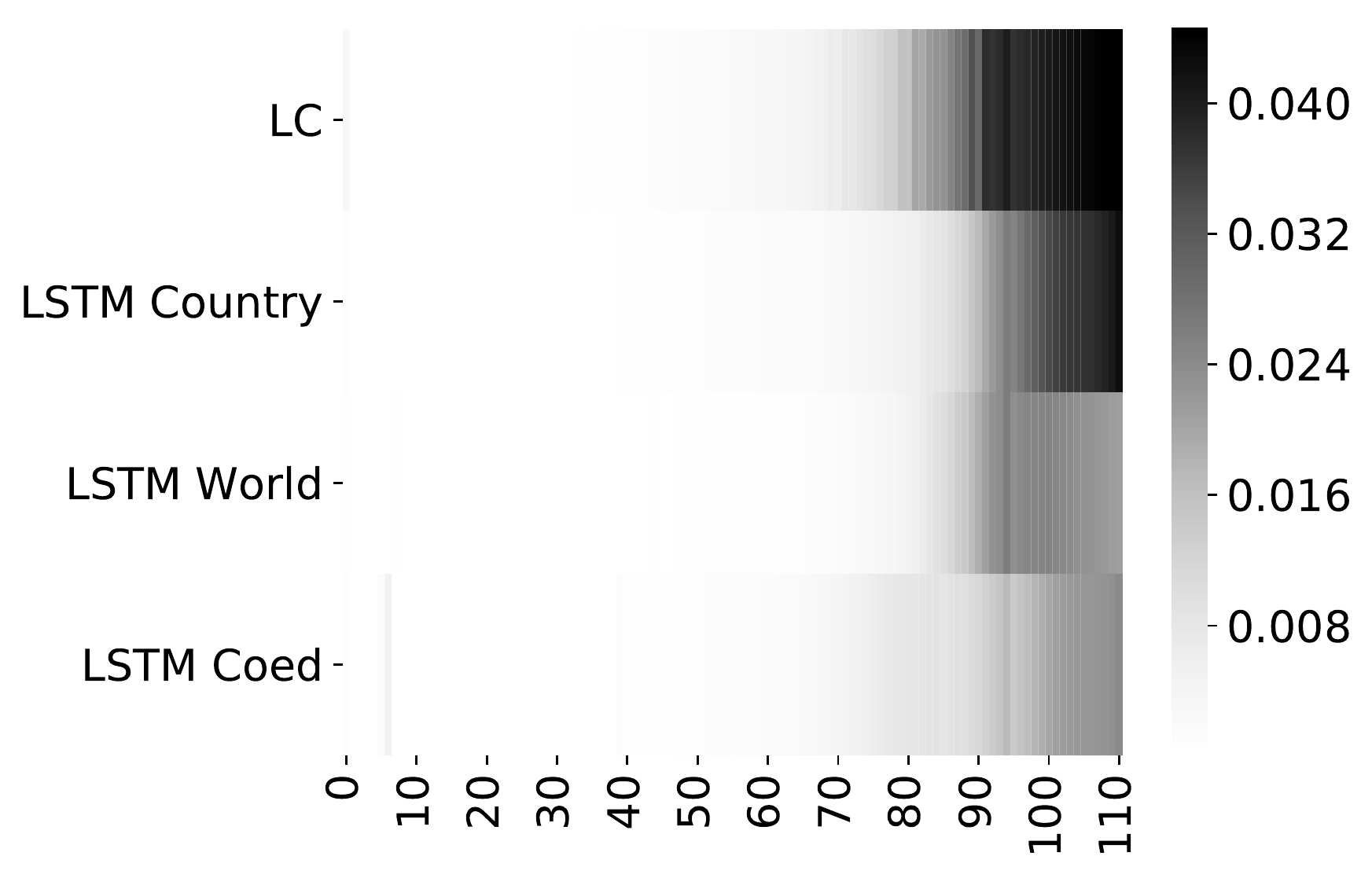}
  \caption{RMSE}
  \label{fig:errors_age_rmse}
\end{subfigure}
\caption{Errors by age}
\label{fig:errors_age}
\end{figure}

Our first LSTM was trained separately for each country, but using the mortality rate series of all ages jointly. It has beaten the LC model in terms of RMSE. However, it produced much higher average SMAPE than the benchmark. The source of this seeming contradiction is that the LSTM struggled in age groups with small mortality, leading to high relative errors. Or rather, it did not struggle at all, just ignored them, since they do not add much to the overall squared error.\\
In order to get rid of this undesirable property, we tried forecasting the natural logarithm of the mortality rates---as does the Lee-Carter model too. We have clipped the values at a lower limit of 1e-12, since some countries' datasets contained zero values.\\
The (standardized) log transformed values produced much more balanced predictions. The forecast quality largely improved in the low-mortality ages, leading to much better relative errors, while keeping the absolute errors nearly at their former levels. The overall RMSE, MAE and MedAE got slightly higher, but it might just be random chance.\\
The LSTM method produced better RMSE than the LC for 27 countries (77\% of all) and for all ages. The SMAPE still tends to be higher for the youth, but it also applies to the Lee-Carter model (Figure \ref{fig:errors_age}). In terms of SMAPE, the LSTM has beaten the LC method for 33 countries (94\%) and for 97 ages (87\%).\\

\begin{figure}
\centering
\includegraphics[width=\textwidth,height=\textheight,keepaspectratio]{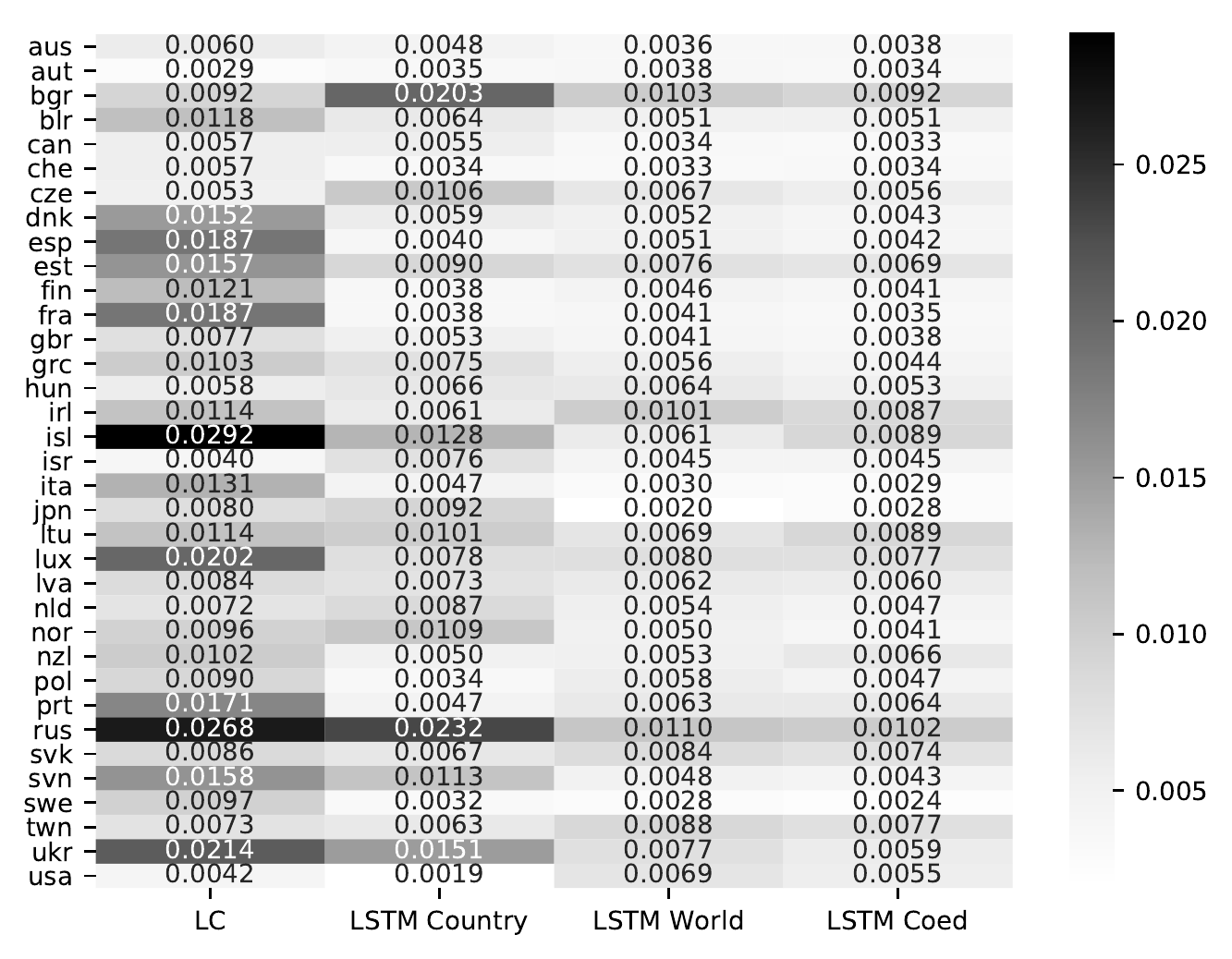}
\caption{RMSE by country}
\label{fig:rmse_country}
\end{figure}

\begin{figure}
\centering
\begin{subfigure}{.5\textwidth}
  \centering
  \includegraphics[width=1.\linewidth]{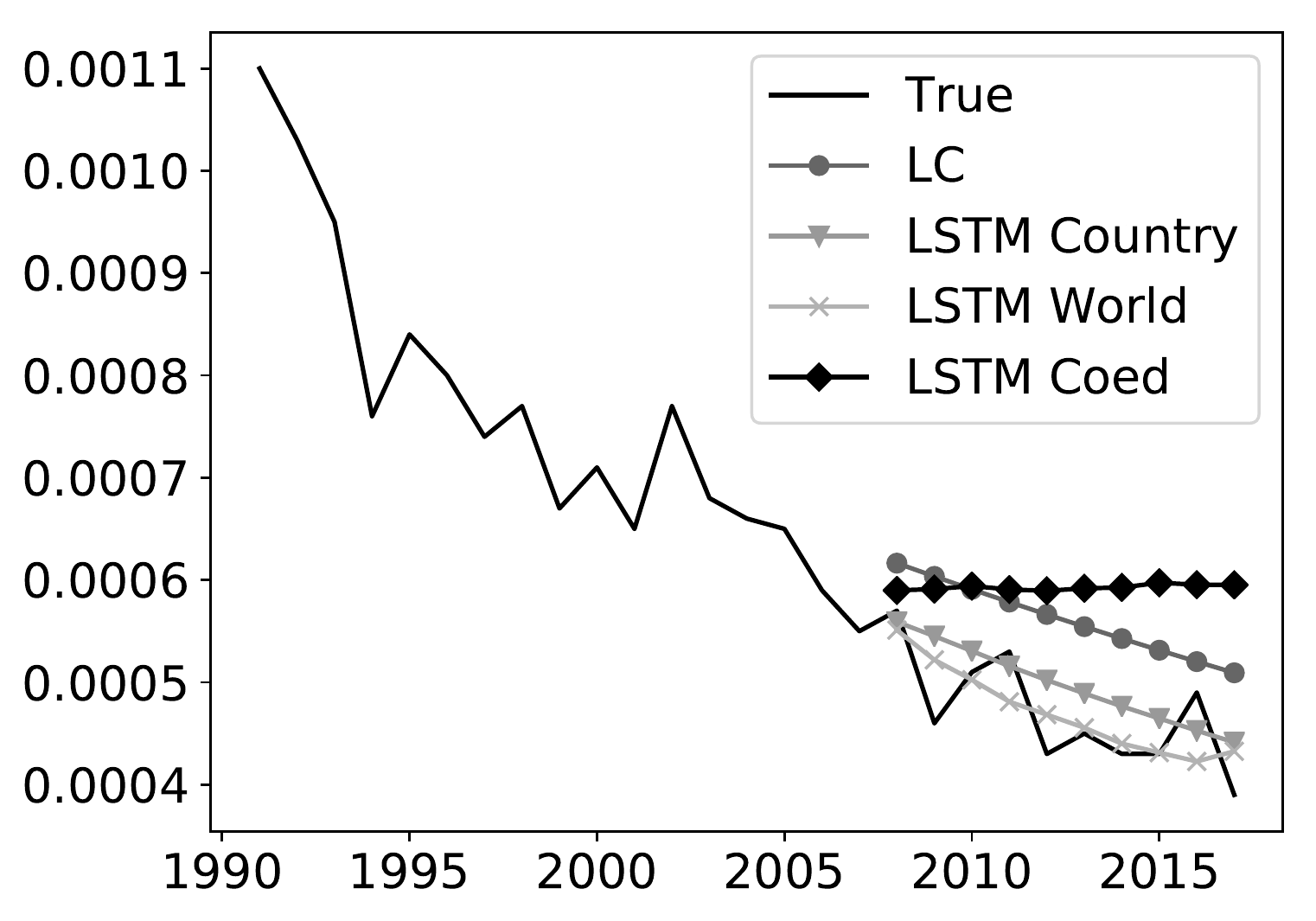}
  \caption{Hungary, Age 25}
  \label{fig:forecasts_hun_25}
\end{subfigure}%
\begin{subfigure}{.5\textwidth}
  \centering
  \includegraphics[width=1.\linewidth]{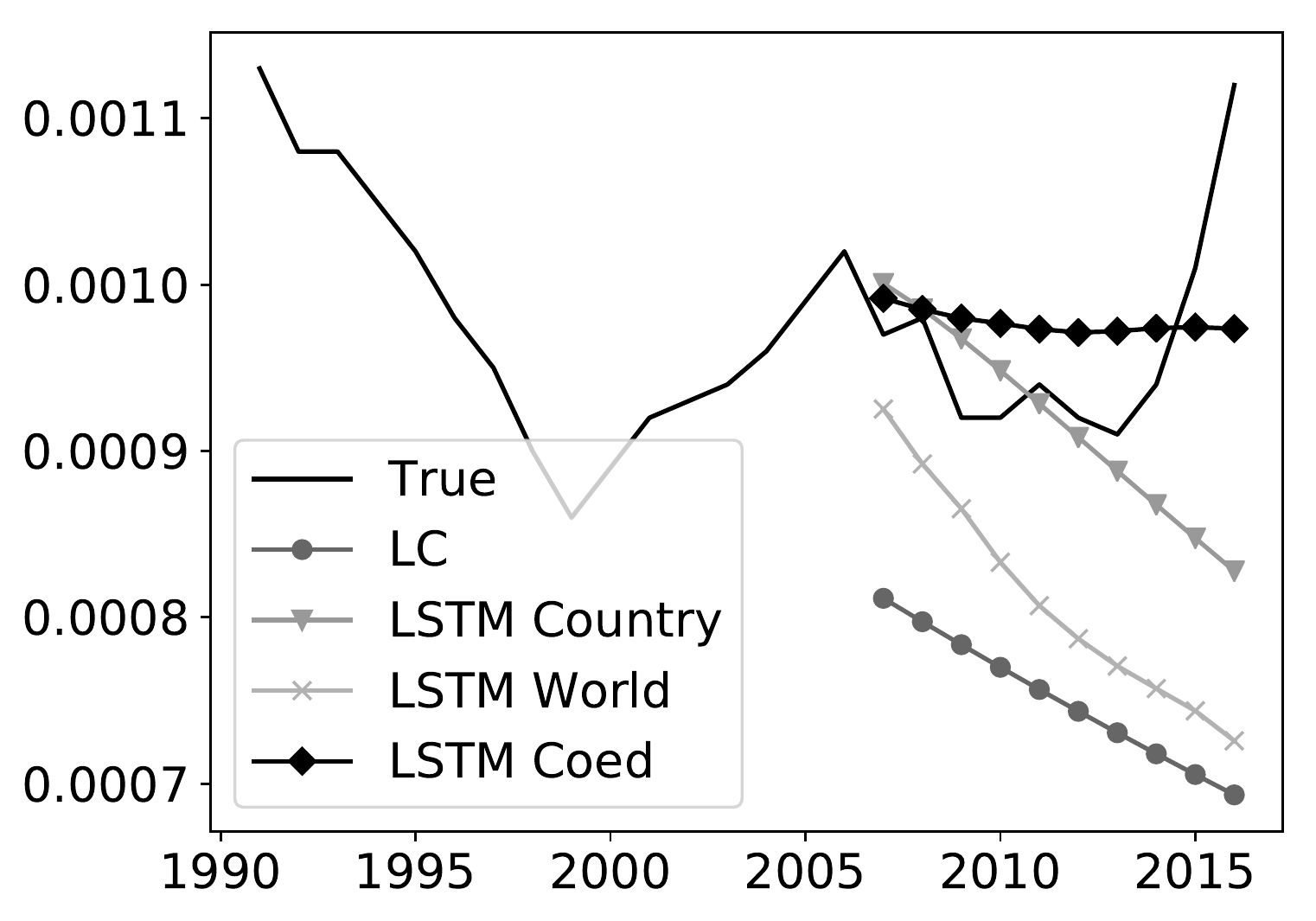}
  \caption{USA, Age 25}
  \label{fig:forecasts_usa_25}
\end{subfigure}
\begin{subfigure}{.5\textwidth}
  \centering
  \includegraphics[width=1.\linewidth]{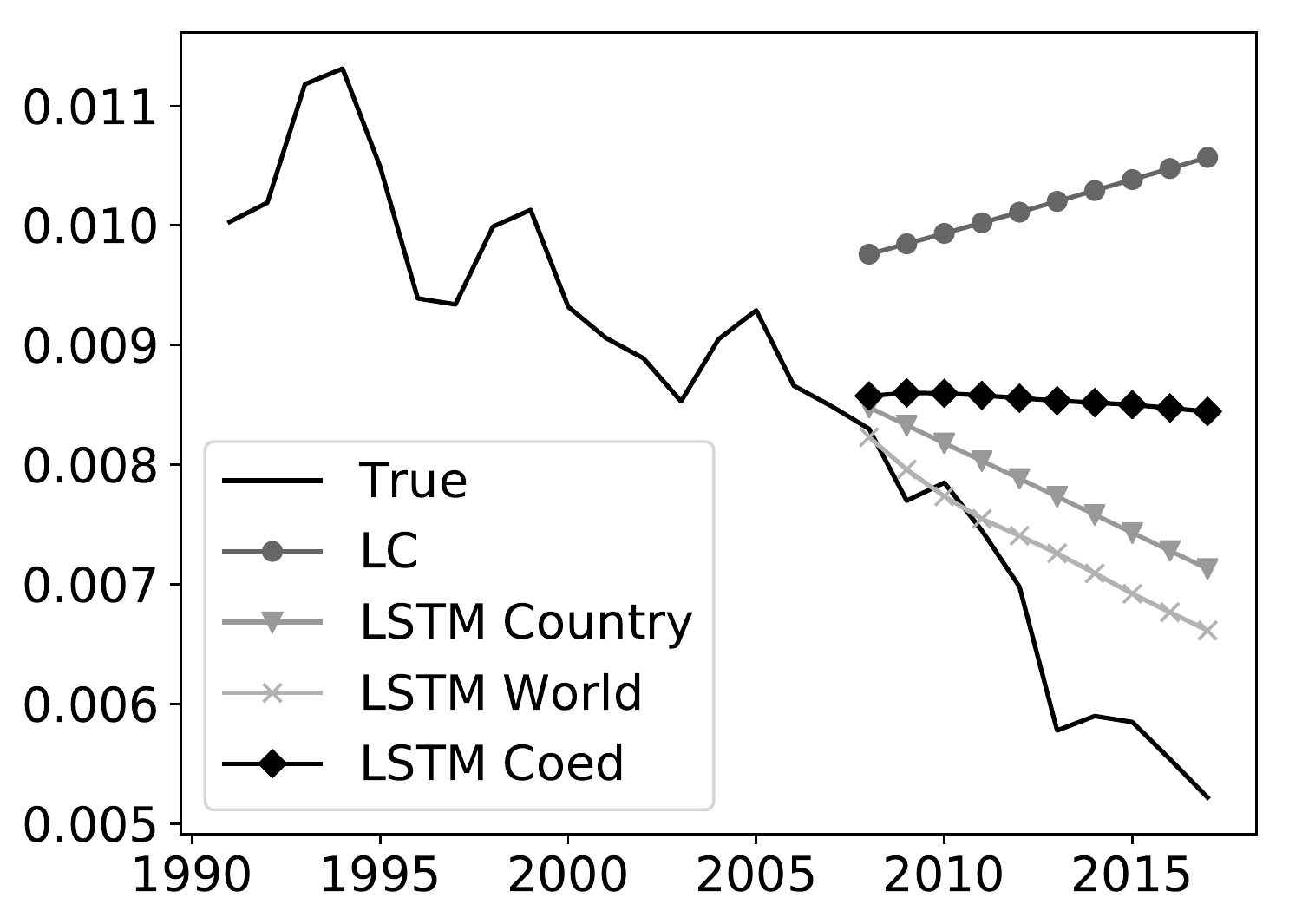}
  \caption{Hungary, Age 50}
  \label{fig:forecasts_hun_50}
\end{subfigure}%
\begin{subfigure}{.5\textwidth}
  \centering
  \includegraphics[width=1.\linewidth]{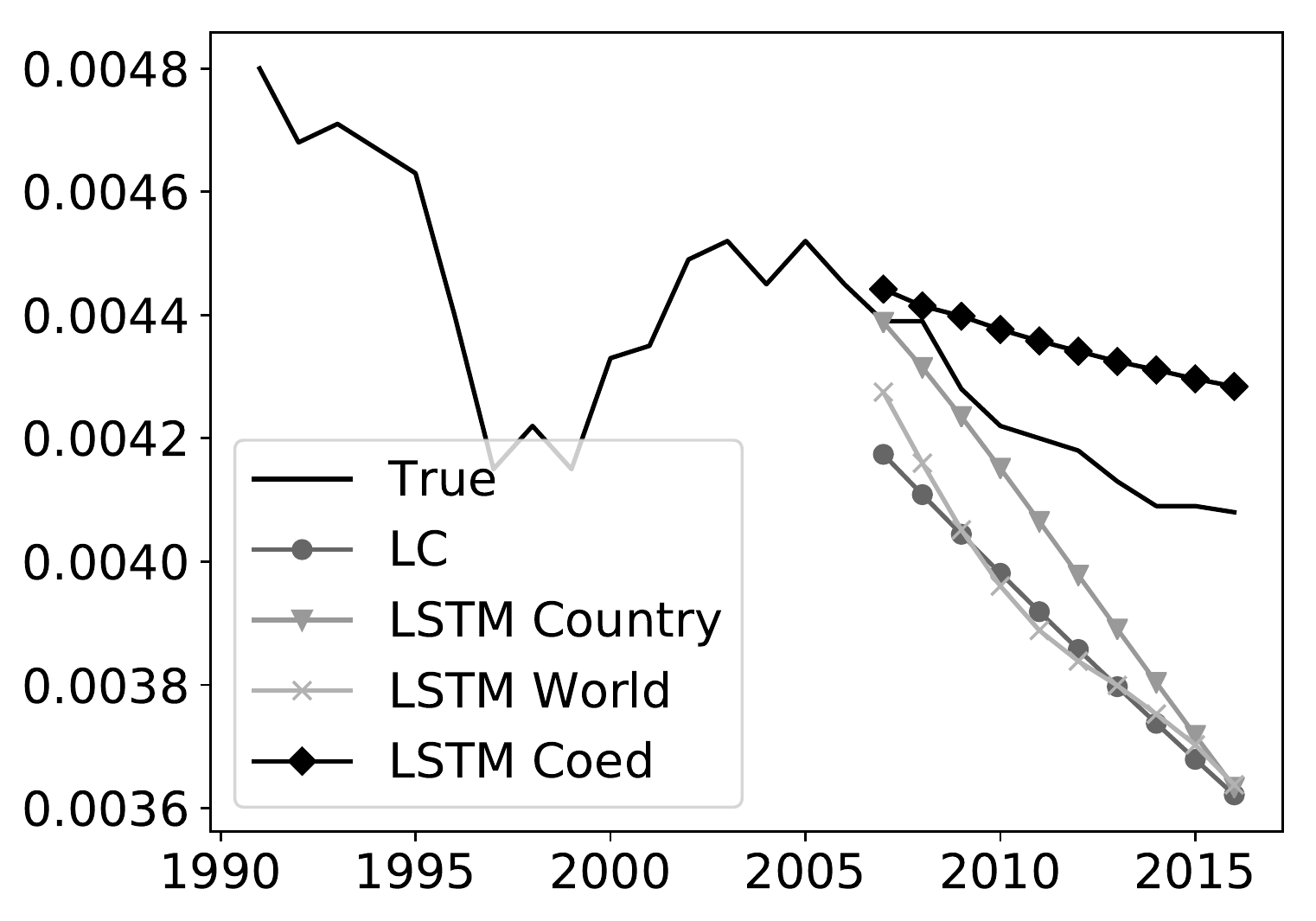}
  \caption{USA, Age 50}
  \label{fig:forecasts_usa_50}
\end{subfigure}
\begin{subfigure}{.5\textwidth}
  \centering
  \includegraphics[width=1.\linewidth]{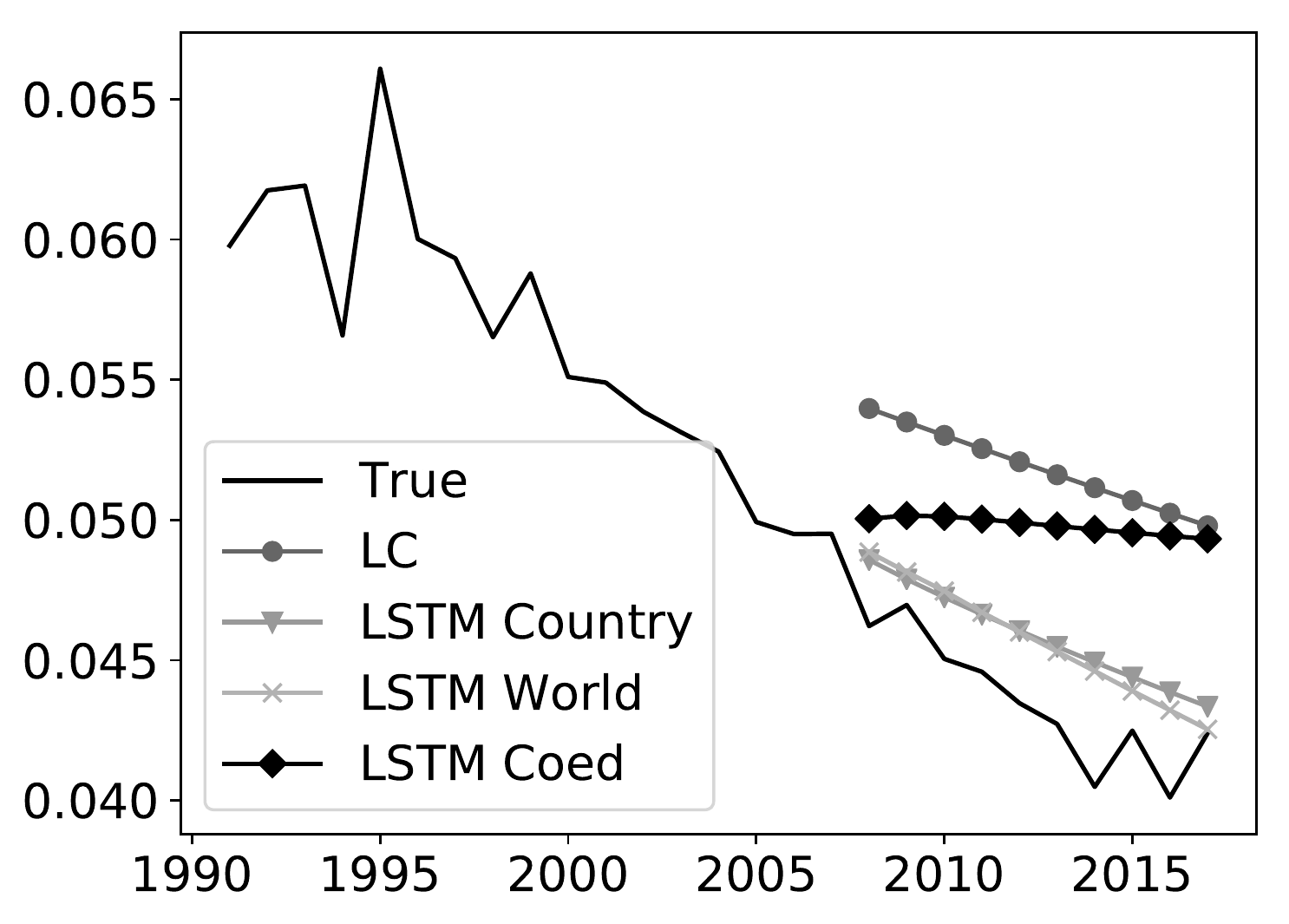}
  \caption{Hungary, Age 75}
  \label{fig:forecasts_hun_75}
\end{subfigure}%
\begin{subfigure}{.5\textwidth}
  \centering
  \includegraphics[width=1.\linewidth]{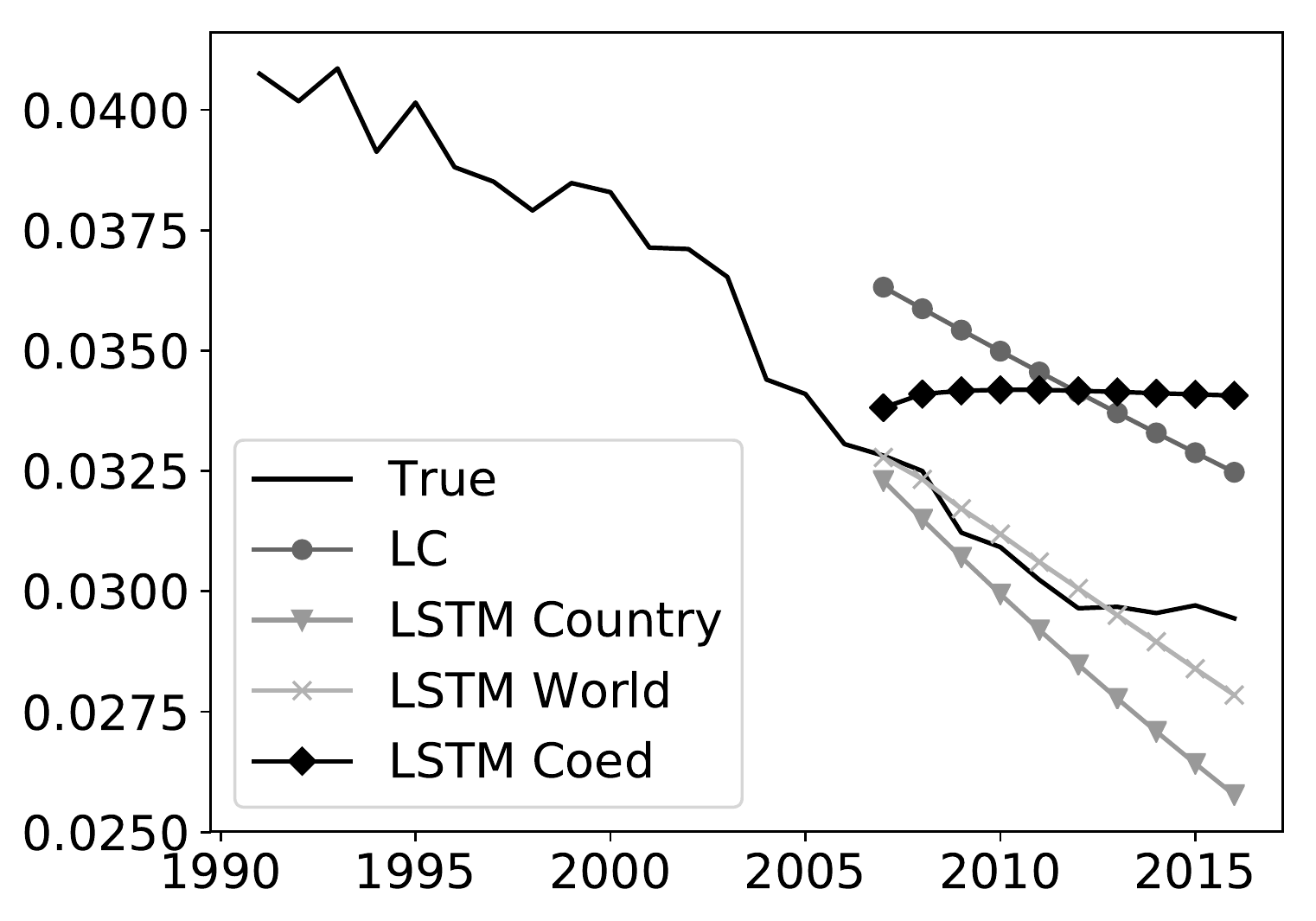}
  \caption{USA, Age 75}
  \label{fig:forecasts_usa_75}
\end{subfigure}
\caption{Forecasts}
\label{fig:forecasts}
\end{figure}

A summary of the overall evaluation results is displayed in Table \ref{table:metrics}. The LSTM performed better on each metric. The mean error is positive for both algorithms, thus they seem to be somewhat positively biased. This bias is higher for the Lee-Carter model.\\
\begin{table}[h!]
\centering
\begin{tabular}{ |c|c|c|c|c| }
 \hline
 metric & LC & LSTM Country & LSTM World & LSTM Coed \\
 \hline
 RMSE&0.0115&0.0076&0.0058&0.0055 \\
 MAE&0.0109&0.0069&0.0051&0.0047 \\
 MedAE&0.0108&0.0067&0.0049&0.0045 \\
 SMAPE&24.85&18.02&15.82&20.76 \\
 ME&0.0085&0.0027&0.0037&0.0026 \\
 \hline
\end{tabular}
\caption{Averaged evaluation metrics}
\label{table:metrics}
\end{table}
The root mean squared errors of the models are shown by country in Figure \ref{fig:rmse_country}.
In case of the LSTM, there’s a negative correlation (-0.25) between the error (RMSE) and length of mortality rate history (years) available for the given country. This measure is slightly positive (0.15) for the Lee-Carter model. It suggests that the LSTM works better for countries with longer recorded mortality rate history. It is not surprising, since neural networks usually shine with lots of data. Thus, irrespective of the results of this Lee-Carter vs. LSTM comparison, it seems that the latter holds more promise for the future. Also, this encouraged us to build LSTMs from all countries' data, in order to improve forecasts for countries with less data. Surprisingly, the negative correlation of errors and history lengths remained, but this approach led to improved forecasts.\\
So, in order to maximize the amount of data used, we trained a single LSTM model from all countries' data. We did not really know what to expect. On the one hand, mortality rates differ from country to country, since they are largely dependent on the local (political, economic, etc.) environment. On the other hand, there still are many similarities that might be exploited, and the LSTM could take advantage of the increased data volume. Also, the LSTM might be flexible enough to learn different mortality patterns, and make forecasts accordingly. To keep it simple, we did not increase the size of the network.\\
This extended model performed better than the Lee-Carter model and the country-level LSTM. It produced better RMSE than the country-level LSTM for 28 countries (80\%), and 103 age groups (93\%).\\
Appetite comes with eating---we grabbed the separate mortality rates of women and men, and fed it to the LSTM. Thus, we tripled our dataset. We still made forecasts only for the total population's mortality rates.\\
This final LSTM has brought further improvement to the overall errors according to most metrics (Table \ref{table:metrics}). In terms of RMSE, it has outperformed even the world-wide LSTM for 26 countries (74\%), however only for 24 ages (22\%). The SMAPE increased compared to the previous LSTM forecasts.\\
Figure \ref{fig:forecasts} displays forecasts for 2 arbitrarily chosen countries in 3 age groups. We can see that the different models typically produce similar forecasts, or at least show a similar direction. However, while the Lee-Carter model spectacularly failed with middle-aged Hungarians' strange mortality rate history, the neural networks produced fairly reasonable forecasts. The Lee-Carter model seemed to work better for the US data, than for most other countries in our study.

\section{Improving the benchmark}
So far, we have applied a fairly simple implementation of the Lee-Carter model. The method might be improved in several ways, which may help keep up with the recurrent neural networks.\\
A possible extension of the Lee-Carter model is using higher order terms from the SVD (\ref{eq:lc_eq_higher}) \citep{booth2001age}. For example, \citet{baran2007forecasting} fitted a third order model to Hungarian mortality data. It may provide a better fit.
\begin{equation} \label{eq:lc_eq_higher}
ln(m_{x,t}) = a_x + b_{1,x}k_{1,t} + b_{2,x}k_{2,t} + ... + b_{n,x}k_{n,t} + e_{x,t}
\end{equation}
The Lee-Carter model was originally applied to US data with a random walk with drift model to forecast the mortality index $k$. However, it may or may not be the best ARIMA model for all countries. Also, when using multiple $k$ series, the right forecasting structure might differ for each. Hence, we should revise the choice of ARIMA parameters.\\
We tried to improve the Lee-Carter model’s forecasting performance by using higher order approximation and automatic time series model selection.\\
We applied a third order Lee-Carter model, and an automatic ARIMA forecasting method proposed by \citet{hyndman2007automatic}: successive KPSS tests for selecting the order of differencing, then a step-wise algorithm to choose the number of AR and MA terms based on the AIC value. The automatically chosen ARIMA parameters are available in Appendix \ref{appendix:a}.\\
The results for the Lee-Carter-extensions are available in Table \ref{table:metrics_extensions} together with the basic Lee-Carter model. The auto ARIMA did not bring much improvement, however the third order model produced remarkably better forecasts. It still seems to underperform the neural network models, but the difference is much smaller.

\begin{table}[h!]
\centering
\begin{tabular}{ |c|c|c|c|c| }
 \hline
 metric & LC & LC Auto & LC Higher & LC Auto Higher \\
 \hline
 RMSE&0.0115&0.0113&0.0078&0.0079 \\
 MAE&0.0109&0.0107&0.0071&0.0072 \\
 MedAE&0.0108&0.0107&0.0070&0.0071 \\
 SMAPE&24.85&24.85&17.28&17.97 \\
 ME&0.0085&0.0083&0.0033&0.0035 \\
 \hline
\end{tabular}
\caption{Averaged evaluation metrics (Lee-Carter variants)}
\label{table:metrics_extensions}
\end{table}

\section{Discussion and Conclusions}
We have compared the 10-year mortality rate forecasting performance of 2 methods: the well-proven Lee-Carter model, and a long short-term memory neural network. The latter is a state-of-the-art method of sequence learning, however it may or may not be successfully applicable to mortality rate forecasting due to data volume issues. Therefore, we built joint neural network models from multiple data series: the mortality rate history of all ages, all countries or even both sexes.\\
Mortality rates of 35 countries were forecasted in 111 age groups. None of the models was fine tuned. We chose to compare simple and general implementations.\\
The results are convincing: the LSTM produced more accurate forecasts. Especially when trained globally and not at the country-level. However, expanding the training set with female and male mortality rate histories did not clearly improve the forecasts of total mortality.\\
This simple comparison has its shortcomings. Our study solely focused on forecasting performance, it did not respect the simplicity and interpretability of the Lee-Carter model, which in itself is valuable. Also, our analysis only considered a single forecasting horizon, and we do not know how the algorithms would perform for the farther future.\\
The proposed LSTM approach could certainly be improved by further expanding the dataset. One possible direction is using data from even more countries. It is clearly limited, since we have already exploited the full depth of the world's most recognized mortality database. Rather, the greatest improvement would be to time travel, at least a few decades, into the future. Time needs to pass for neural networks to occupy their worthy place in mortality rate forecasting.

\bibliography{bib}

\clearpage

\appendix
\section{Appendix: Automatic ARIMA parameters}
\label{appendix:a}

\begin{table}[h!]
\centering
\begin{tabular}{ |c|ccc|ccc|ccc|ccc| }
 \hline
 \multirow{3}{*}{country} & \multicolumn{3}{c|}{LC} & \multicolumn{9}{c|}{LC Higher} \\
 \cline{2-13}
  & \multicolumn{3}{c|}{$k$} & \multicolumn{3}{c}{$k_1$} & \multicolumn{3}{c}{$k_2$} & \multicolumn{3}{c|}{$k_3$} \\
 \cline{2-13}
  & AR & I & MA & AR & I & MA & AR & I & MA & AR & I & MA \\
 \hline
aus & 0 & 1 & 1 & 0 & 1 & 1 & 1 & 2 & 1 & 1 & 0 & 0 \\
aut & 0 & 1 & 0 & 0 & 1 & 0 & 0 & 2 & 2 & 1 & 0 & 0 \\
bgr & 0 & 2 & 3 & 0 & 2 & 3 & 0 & 1 & 1 & 1 & 0 & 0 \\
blr & 0 & 1 & 0 & 0 & 1 & 0 & 1 & 0 & 0 & 1 & 0 & 0 \\
can & 0 & 1 & 0 & 0 & 1 & 0 & 1 & 2 & 1 & 2 & 0 & 0 \\
che & 0 & 1 & 1 & 0 & 1 & 1 & 0 & 1 & 1 & 1 & 1 & 3 \\
cze & 0 & 1 & 0 & 0 & 1 & 0 & 2 & 2 & 1 & 1 & 0 & 0 \\
dnk & 1 & 1 & 1 & 1 & 1 & 1 & 0 & 1 & 1 & 1 & 1 & 1 \\
esp & 1 & 1 & 0 & 1 & 1 & 0 & 0 & 1 & 1 & 1 & 0 & 0 \\
est & 0 & 1 & 3 & 0 & 1 & 3 & 1 & 1 & 0 & 0 & 1 & 0 \\
fin & 0 & 1 & 0 & 0 & 1 & 0 & 0 & 1 & 1 & 1 & 1 & 1 \\
fra & 1 & 1 & 1 & 1 & 1 & 1 & 1 & 1 & 1 & 0 & 1 & 1 \\
gbr & 2 & 1 & 2 & 2 & 1 & 2 & 1 & 2 & 1 & 2 & 0 & 2 \\
grc & 1 & 1 & 0 & 1 & 1 & 0 & 0 & 0 & 0 & 0 & 0 & 0 \\
hun & 1 & 1 & 2 & 1 & 1 & 2 & 1 & 1 & 1 & 1 & 2 & 1 \\
irl & 2 & 1 & 0 & 2 & 1 & 0 & 0 & 2 & 1 & 2 & 0 & 0 \\
isl & 2 & 1 & 0 & 2 & 1 & 0 & 2 & 1 & 1 & 0 & 0 & 0 \\
isr & 0 & 1 & 0 & 0 & 1 & 0 & 0 & 0 & 0 & 0 & 0 & 0 \\
ita & 0 & 1 & 0 & 0 & 1 & 0 & 3 & 0 & 2 & 1 & 1 & 1 \\
jpn & 0 & 2 & 3 & 0 & 2 & 3 & 2 & 2 & 1 & 2 & 0 & 1 \\
ltu & 0 & 1 & 1 & 0 & 1 & 1 & 2 & 0 & 1 & 1 & 0 & 1 \\
lux & 2 & 1 & 0 & 2 & 1 & 0 & 1 & 2 & 1 & 0 & 0 & 0 \\
lva & 0 & 1 & 0 & 0 & 1 & 0 & 2 & 0 & 0 & 3 & 0 & 2 \\
nld & 1 & 1 & 1 & 1 & 1 & 1 & 0 & 1 & 2 & 0 & 1 & 1 \\
nor & 1 & 1 & 0 & 1 & 1 & 0 & 0 & 1 & 1 & 0 & 1 & 1 \\
nzl & 1 & 1 & 0 & 1 & 1 & 0 & 1 & 1 & 0 & 2 & 0 & 0 \\
pol & 1 & 1 & 0 & 1 & 1 & 0 & 0 & 1 & 0 & 2 & 0 & 0 \\
prt & 2 & 1 & 2 & 2 & 1 & 2 & 0 & 2 & 1 & 1 & 1 & 0 \\
rus & 0 & 1 & 1 & 0 & 1 & 1 & 1 & 0 & 2 & 1 & 0 & 0 \\
svk & 0 & 1 & 0 & 0 & 1 & 0 & 0 & 1 & 0 & 0 & 0 & 0 \\
svn & 0 & 1 & 0 & 0 & 1 & 0 & 0 & 1 & 0 & 0 & 0 & 0 \\
swe & 1 & 1 & 2 & 1 & 1 & 2 & 3 & 1 & 3 & 2 & 1 & 1 \\
twn & 0 & 1 & 0 & 0 & 1 & 0 & 1 & 0 & 0 & 1 & 0 & 0 \\
ukr & 0 & 1 & 0 & 0 & 1 & 0 & 2 & 0 & 1 & 1 & 0 & 0 \\
usa & 1 & 1 & 0 & 1 & 1 & 0 & 0 & 2 & 1 & 2 & 1 & 2 \\
 \hline
\end{tabular}
\caption{Auto ARIMA parameters}
\label{table:metrics_extensions}
\end{table}

\end{document}